# EUFormer: Learning Driven 3D Spine Deformity Assessment with Orthogonal Optical Images

Nan Meng[1,2,3,†], *IEEE Member*, Jason P.Y. Cheung[1,3], Tao Huang[1,2,3,†], Moxin Zhao[1,2], Yue Zhang[1,2], Chenxi Yu[1,2], Chang Shi[1,2], Teng Zhang[1,2,3,*], *IEEE Senior Member*

[1] Department of Orthopaedics and Traumatology, The University of Hong Kong
[2] Digtal Health Laboratory, The University of Hong Kong
[3] Conova Medical Technology Limited
{nanmeng, cheungjp, huangtao, moxin, yilzhang, ychxhku, chaseshi, tgzhang}@hku.hk

*Abstract*—In clinical settings, the screening, diagnosis, and monitoring of adolescent idiopathic scoliosis (AIS) typically involve physical or radiographic examinations. However, physical examinations are subjective, while radiographic examinations expose patients to harmful radiation. Consequently, we propose a pipeline that can accurately determine scoliosis severity. This pipeline utilizes posteroanterior (PA) and lateral (LAT) RGB images as input to generate spine curve maps, which are then used to reconstruct the three-dimensional (3D) spine curve for AIS severity grading. To generate the 2D spine curves accurately and efficiently, we further propose an Efficient U-shape transFormer (EUFormer) as the generator. It can efficiently utilize the learned feature across channels, therefore producing consecutive spine curves from both PA and LAT views. Experimental results demonstrate superior performance of EUFormer on spine curve generation against other classical U-shape models. This finding demonstrates that the proposed method for grading the severity of AIS, based on a 3D spine curve, is more accurate when compared to using a 2D spine curve.

## I. INTRODUCTION

Scoliosis is an orthopaedic disorder characterized by an abnormal lateral curvature of the spine. The most common type of scoliosis is adolescent idiopathic scoliosis (AIS), which affects approximately 2-3% of adolescents [1]. AIS generally presents around the age of 10, with the potential for rapid progression during puberty [2]. The diagnosis of scoliosis is often overlooked or delayed since most patients in the early stages do not experience symptoms. However, some patients may develop complications such as impaired cardiopulmonary function, chronic back pain, and body image disturbances [3]. It is crucial to conduct AIS screening for early diagnosis and ensure close follow-up after diagnosis in order to tailor intervention strategies and mitigate the risk of further progression [4, 5]. Clinically, physical examination, or Cobb angle (CA) acquired from radiographs are used to assess the severity of AIS, however, physical examination results are subjective leading to inaccurate inference, while radiography expose harmful radiation to patients [6]. Therefore, developing a radiation-free approach to accurately assess scoliosis severity is of paramount importance for effective screening and monitoring of the condition.

In order to achieve AIS assessment without radiation exposure, optical images have been considered as a promising alternative. Zhang et al. [7] proposed utilizing smartphone photos of bare backs to train a classification network for categorizing severity levels. However, the classification model provides severity directly, rendering it a black box, with the internal processes remaining unknown. To extract the decisive factor, namely the spine shape, from optical images, Meng et al. [8] proposed utilizing back view RGB-Depth images to create artificial spine X-rays for assessing scoliosis severity. However, direct synthesis of X-rays presents significant challenges since the input images are sensitive to the shooting environment, resulting in poor generation outcomes. In this study, we explore the generation of 2D spine curve maps from orthogonal optical images of the unclothed trunk. This approach offers a more concise, robust, and practical solution compared to X-ray synthesis.

The spine curve generation process employs a generative adversarial network (GAN) model. To effectively utilize the information from patients' unclothed trunks, we propose an Efficient U-shape transFormer (EUFormer) as the generator. EUFormer comprises multiple efficient transformer blocks (ETB). Unlike the standard vision transformer (ViT) [9], each ETB conducts attention in the channel dimension rather than the spatial dimensions. This strategy significantly reduces the computations required for self-attention operation. Furthermore, a revised LeFF module is adopted to replace the MLP layer in ViT, as it can efficiently leverage local context. Upon obtaining the two orthogonal spine curve maps, the 3D spine curve can be reconstructed based on them. We subsequently propose a method to identify the severity level using the 3D spine curve, which demonstrates superior performance compared to the 2D spine curve approach.

The contributions of our study include:

1. We propose a radiation-free and feasible pipeline for scoliosis severity grading based only on optical RGB images.
2. We propose the ETB module and base on such module we construct the EUFormer which can efficiently utilize the learned feature maps of patient's unclothed trunk across various channels, thereby producing high-quality and consecutive spine curve maps from RGB images.
3. We explore a novel scoliosis severity grading method based on 3D spine curve, with experimental results indicating enhanced performance.

---

[*] Corresponding author
[†] Contributed equally to this work.

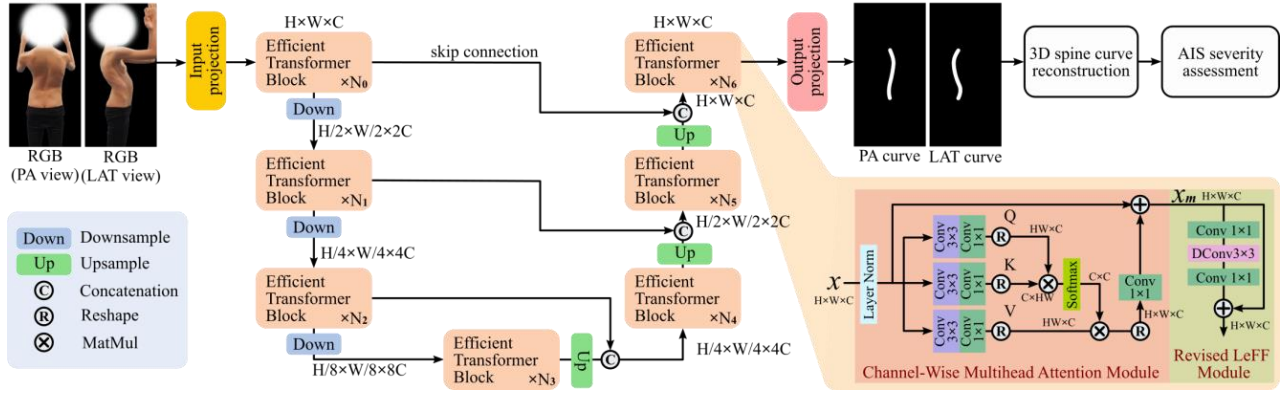

**Fig. 1.** Overview of the proposed pipeline for posteroanterior (PA) and lateral (LAT) spine curve synthesis and 3D spine curve reconstruction. The basic component is Efficient Transformer Block (ETB) which consists of a channel-wise multihead attention module and a revised LeFF module.

## II. METHOD

### A. Dataset

This study recruited patients with AIS from two territory-wide tertiary scoliosis referral centers (72% female; age range 10-18). The study was approved by the local institutional authority review board (UW15-596). Written informed consent was obtained from all participants. Exclusion criteria included psychological or systematic neural disorders, congenital deformities, previous spinal operations, trauma impairing posture and mobility, and oncological diseases. A total of 451 patients were analyzed (67% female; age range 10-18). Each patient was required to capture a posteroanterior (PA) RGB image and a lateral (LAT) RGB image of unclothed trunk and biplanar X-rays of spine. The dataset was divided into two subsets: a training set and a test set. The training set included 376 cases, while the test set comprised 75 prospective cases not used during the training phase. Spine specialists physically located in the spine clinics measured the ground truth value of CA using the picture archiving and communication system.

The ground truth (GT) of the orthogonal spine curve maps were obtained from biplanar X-rays. We first align the PA and LAT X-rays with corresponding RGB images. The alignment was based on landmarks annotated by our spine experts. Then, the region of spine in both biplanar X-rays was segmented out using our previous model [10], resulting in two orthogonal spine curve maps.

### B. Method Pipeline

The proposed pipeline for AIS severity assessment is depicted in Fig. 1. It comprises an EUFormer for generating PA and LAT spine curve maps, followed by a 3D spine curve reconstruction module to reconstruct the 3D spine curve. Lastly, an AIS severity identification module is employed to infer the severity level according to the 3D spine curve.

*1) Efficient U-shape Transformer (EUFormer)*

The primary computational burden in Transformers stems from the self-attention layer where the time and memory complexity of the key-query dot product interaction escalates quadratically in relation to the input's spatial resolution [9, 11]. To reduce the computations, we introduce an efficient attention method that applies self-attention in the channel dimension. This method facilitates the integration of the learned features across different feature channels, thereby benefiting the generation of high-quality spine curve. Fig. 1 presents the detailed structure of EUFormer. As shown, the EUFormer is a U-shaped model that comprises several efficient transformer blocks (ETBs). Each ETB consists of a *channel-wise multi-head attention (CMHA)* module and a *revised LeFF* module [12].

Within the CMHA module, given an input tensor $x \in \mathbb{R}^{H \times W \times C}$, the module first computes query (Q), key (K) and value (V) projections. This is accomplished by employing a 3×3 convolution for spatial feature extraction, as well as a 1×1 convolution to aggregate pixel-wise cross-channel features, resulting in $Q = W_m^Q W_n^Q x$, $K = W_m^K W_n^K x$, $V = W_m^V W_n^V x$. $W_m^{(\cdot)}$ and $W_n^{(\cdot)}$ are the weights for 3×3 convolution and 1×1 convolution, respectively. In our model, we employ convolutional layers without bias. Next, we reshape the Q and K matrices and fuse their spatial dimensions (H and W), yielding a channel attention map A of size C×C, in contrast to the traditional attention map of size HW×HW [9]. The entire CMHA process can be formulated as follows:

$$x_m = V \cdot \text{Softmax}(K \cdot Q/\alpha) + x, \quad (1)$$

where $x$ and $x_m$ are the input and output feature maps of the CMHA module. $Q \in \mathbb{R}^{HW \times C}$, $K \in \mathbb{R}^{C \times HW}$, and $V \in \mathbb{R}^{HW \times C}$ are matrices after reshape operation. $\alpha$ serves as a learnable scaling factor that regulates the magnitude of the dot product between K and Q prior to implementing the Softmax function. Similar to the traditional multi-head self-attention (SA) [9], we partition the number of channels into 'heads' and concurrently learn distinct attention maps.

The outputs of CMHA module were then fed into a revised LeFF module. As suggested by Wang et al. [12], the LeFF module can efficiently utilize local context from adjacent pixels which plays a critical role in dense prediction problem. Here, to fit for our model pipeline, we revised the LeFF module by removing two reshape operations between the convolutional layers with in LeFF module. In this way, the entire ETB module can preserve the shape of input feature maps and therefore enabling feature concatenation (skip connection) between features with the same spatial scales in the U-shape architecture, as shown in Fig. 1.

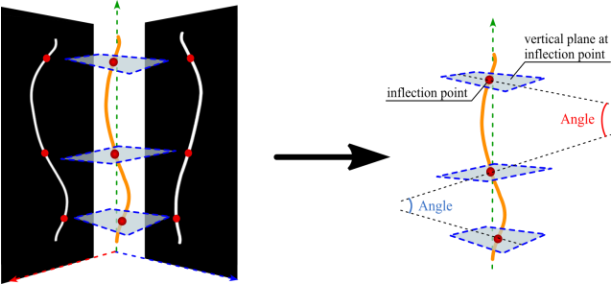

**Fig. 2.** The diagram of 3D spine curve reconstruction process and the 3D Cobb angle calculation.

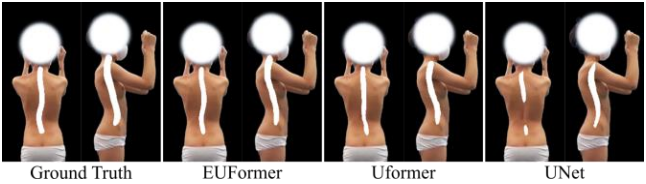

**Fig. 3.** Visual comparison of EUFormer against Uformer and UNet on orthogonal spine curve generation.

The discriminator used in the proposed model was the same as the 5-layer PatchGAN [13]. Both downsampling and upsampling operations were implemented utilizing a pixel-shuffle layer [14] with a convolutional layer.

### C. 3D Spine Curve Reconstruction and AIS severity Classification

The PA and LAT spine curve maps generated by our proposed EUFormer are subsequently employed to calculate the 3D spine curve according to the orthogonal relationship between the two curve maps.

Following this, the position of each end vertebra is pinpointed as the inflection point within the 3D curve, in accordance with the method suggested by [15]. In contrast to the traditional in-plane CA calculation, our research incorporates the 3D spatial configuration of the spine. To accomplish this, planes perpendicular (normal) to the spinal curve at the inflection points are constructed, and the intersections of these planes are used to ascertain the largest angle between them. This angle serves as a criterion for disease severity classification, as depicted in Fig. 2. For convenience, in this paper, we call the angle between the two vertical planes at adjacent inflection points the "3D CA". The value of a 3D CA less than 20° are considered normal-mild, those ranging from 20° to 40° are deemed moderate, and values exceeding 40° are classified as severe.

## III. EXPERIMENTAL DETAILS

### A. Experimental Implementation

The proposed model has been developed using the PyTorch framework and optimized with the ADAM solver on an NVIDIA RTX 3090 GPU. An initial learning rate of 1e-4 is employed and subsequently reduced by a factor of 0.1 after every 50 epochs. To enhance the training process, input images undergo augmentation through random rotation and flipping. Both generator and discriminator were trained for 100 epochs. Each pair of orthogonal RGB images are resized to 320×160 before fed into the generator.

TABLE I. QUANTITATIVE EVALUATION OF SEVERITY GRADING

| Evaluation metrics | Severity level | | |
|---|---|---|---|
| | Normal-Mild | Moderate | Severe |
| Sensitivity | 0.915 | 0.857 | 0.857 |
| Specificity | 0.893 | 0.907 | 1.000 |
| Precision | 0.935 | 0.783 | 1.000 |
| NPV | 0.862 | 0.942 | 0.986 |
| Accuracy | 0.907 | 0.893 | 0.987 |

**NPV**: negative prediction value.

### B. Loss Function

The proposed EUFormer generator was optimized using a generative loss item $\ell_g$ and an MSE loss item $\ell_2$. The generative loss was used to drive our generator to find solutions that reside on the manifold of spine curve maps, which can be formulated as:

$$\ell_g = \sum_{n=1}^{N} -[y_n \cdot \log(\kappa_n) + (1-y_n) \cdot \log(1-\kappa_n)], \quad (2)$$

where $\kappa_n$ is the probability obtained from the discriminator D when fed the $n^{th}$ generated spine curve map and $y_n$ denotes the $n^{th}$ binary label. The MSE loss encourages the generator to produce realistic spine curve which can be formulated as:

$$\ell_2 = \frac{1}{N}\sum_{n=1}^{N}\left(I_n^S - G_{\theta_G}(I_n^{RGB})\right)^2, \quad (3)$$

where $I_n^S$ denotes the $n^{th}$ ground truth spine curve map while $I_n^{RGB}$ represents the $n^{th}$ RGB input image. $G_{\theta_G}$ denotes the generator with a trainable parameter set $\theta_G$. As a result, the entire training loss for optimizing the generator can be expressed as:

$$\mathcal{L} = \alpha\ell_g + \ell_2, \quad (4)$$

where α is the hyper-parameter weighting for generative loss item with the value of 0.01.

## IV. RESULTS AND DISCUSSION

### A. Spine Curve Generation

The direct outputs of the proposed model are orthogonal (PA and LAT) spine curve maps, as shown in Fig. 1. The precision of the generated spine curve is closely related to the severity grading of the AIS. To validate the model effectiveness, we compared the performance of the proposed EUFormer generator against two classic U-shape model, i.e., UNet [16] and Uformer [12]. UNet is a fully convolutional network with restricted receptive field while Uformer is a ViT-based framework using traditional spatial self-attention mechanisms. Table 1 presents the results in terms of IoU and dice similarity coefficient. For both metrics, the proposed model achieves the highest value, demonstrating superior performance on spine curve generation. Fig. 3 visualizes the spine curve generation results. The proposed EUFormer can achieve high-quality and consecutive spine curves compared against Uformer and UNet.

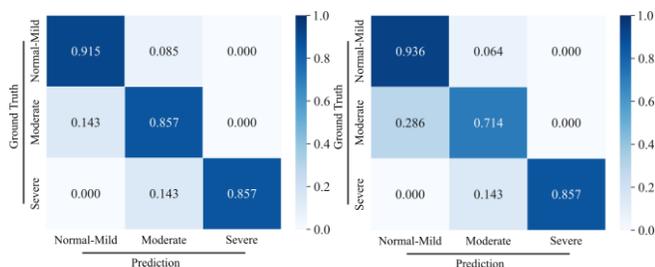

**Fig. 4.** Confusion matrices on severity grading using 3D spine curve (left) and 2D spine curve (right).

TABLE II. QUANTITATIVE COMPARISON ON SPINE CURVE GENERATION

|  | IoU↑ | Dice similarity↑ | Macro Avg. Sensitivity |
|---|---|---|---|
| UNet [16] | 52.3 | 0.68 | 0.793 |
| Uformer [12] | 54.2 | 0.69 | 0.812 |
| EUFormer | **56.3** | **0.71** | **0.876** |

**SD**: standard deviation

Although the dataset predominantly consists of female participants, the model demonstrates consistent performance in analyzing data from both male and female patients, with no significant differences observed.

*B. Severity grading*

We concurrently validated the efficacy of utilizing the 3D spine curve, derived from the orthogonal spine curve maps produced by the model, for severity grading purposes of AIS. Compared with traditional CA, the proposed 3D CA validation pipeline fully considers the morphology of spine in 3D space. The ground truth of scoliosis severity level for each patient was determined based on the X-ray. Fig. 4 compares the performance of severity grading using 3D spine curve (left) and 2D spine curve (right). The two methods show competitive performance when assessing severe cases. This can primarily be attributed to the fact that, in severe patients, the surface features of the unclothed trunk are already obvious, and the 2D spine curve is adequate for distinguishing the severe cases. The classification outcomes using 3D CA are superior for moderate cases compared to the results acquired through 2D spine curve (PA view) analysis, while for normal-mild cases, the results acquired by 2D spine curve are better. For patients with normal-mild or moderate conditions, the surface features of the trunk are less pronounced, causing the curvature of the spine curve to be underestimated. Consequently, the diagnostic accuracy for moderate patients based on the 2D spine curve method is diminished which also results in an inflated sensitivity index for normal-mild cases.

Table 2 displays a more comprehensive evaluation of using 3D CA for severity grading. The proposed pipeline achieves a high degree of sensitivity and NPV when evaluating disease severity across all three levels (*normal-mild:* sensitivity=0.915, NPV=0.862; *moderate:* sensitivity=0.857, NPV=0.942; severe: sensitivity=0.857, NPV=0.986). These findings substantiate the effectiveness of our proposed approach for assessing AIS disease severity, utilizing the 3D CA method.

## V. CONCLUSION

In this study, we propose a radiation-free pipeline that reconstructs the 3D spine curve for severity grading in AIS. The reconstruction of 3D spine curve is based on two orthogonal 2D spine curve maps. To reconstruct high-quality spine curve maps, we propose EUFormer, an efficient U-shape transformer that can utilize information across different channels of feature maps. Experimental results show that the pipeline proposed in this study shows satisfied sensitivity and NPV in severity grading for AIS. The finding proves that the proposed pipeline offers a precise and feasible solution for non-radiation AIS screening and monitoring.

ACKNOWLEDGMENT

This work was supported by Department Seed Fund of The University of Hong Kong (sponsor: Dr. Nan Meng), Innovation and Technology Fund (MRP/038/20X) and Health Services Research Fund (HMRF) 08192266.